\documentclass[aps,preprint,balancelastpage,nofootinbib,12pt,groupedaddress,onecolumn]{revtex4}
\usepackage{amsfonts}
\usepackage{amsmath}
\usepackage{amssymb}
\usepackage[dvips]{graphicx}
\usepackage{epsfig}

\providecommand{\U}[1]{\protect\rule{.1in}{.1in}}

\begin{document}

\title{Logarithmic behavior of degradation dynamics in metal--oxide
semiconductor devices }
\author{Roberto da Silva}
\affiliation{Instituto de Inform\'{a}tica - UFRGS}
\author{Gilson I. Wirth}
\affiliation{Escola de Engenharia - UFRGS}

\begin{abstract}
In this paper the authors describe a theoretical simple statistical
modelling of relaxation process in metal-oxide semiconductor devices that
governs its degradation. Basically, starting from an initial state where a
given number of traps are occupied, the dynamics of the relaxation process
is measured calculating the density of occupied traps and its fluctuations
(second moment) as function of time. Our theoretical results show a
universal logarithmic law for the density of occupied traps $\overline{%
\left\langle n(t)\right\rangle }\ \sim \varphi (T,E_{F})\ (A+B\ \ln t)$,
i.e., the degradation is logarithmic and its amplitude depends on the
temperature and Fermi Level of device. Our approach reduces the work to the
averages determined by simple binomial sums that are corroborated by our
Monte Carlo simulations and by experimental results from literature \cite%
{Tibor2009} which bear in mind enlightening elucidations about the physics
of degradation of semiconductor devices of our modern life.
\end{abstract}

\maketitle

The understanding of the physics of semiconductor devices \cite{Kirton1989}
has never been so important, since silicon-based integrated circuits are
facing increasing reliability and scaling issues. Quantum Computing \cite%
{Feynman1982}, DNA Computing \cite{Adleman1994}, and many other alternatives
are arising as possible substitutes to the technology of the silicon-based
computation, but these seem to be very distant from the reality of our
day-by-day.

Hence, the understanding of the reliability effects in semiconductors is of
paramount importance for the Microelectronics Industry, that is more and
more dependent of new results from an interdisciplinary Physics and their
ramifications.

\begin{figure}[tbp]
\begin{center}
\includegraphics[width=0.9\columnwidth]{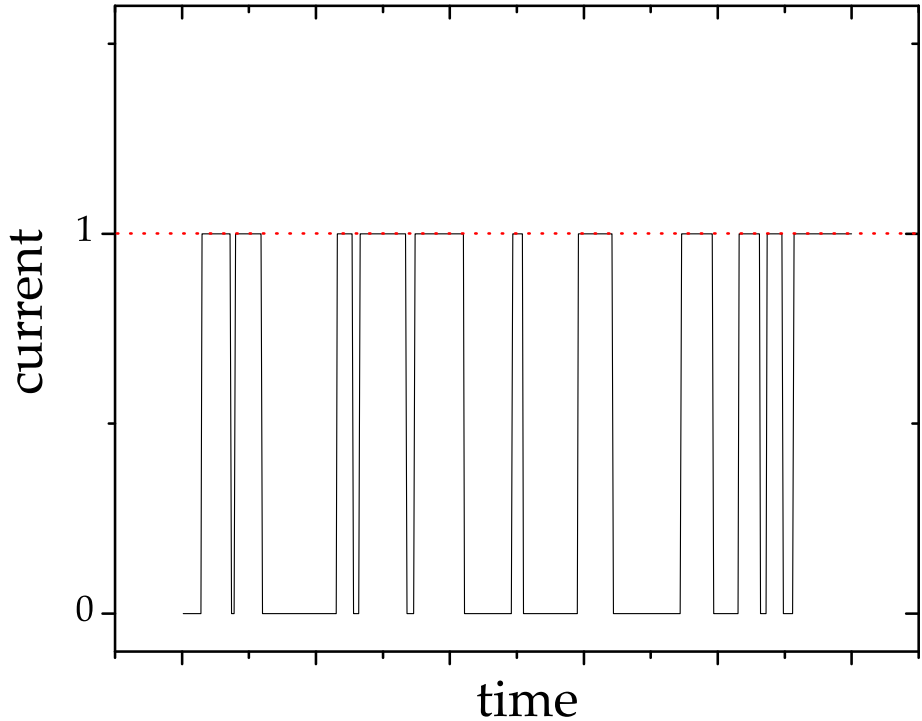}
\end{center}
\caption{A simple representation of a random telegraph signal caused by
sucessive captures and emissions of charge carriers by a trap in the current
of a CMOS transistor. }
\label{rts}
\end{figure}

Complementary metal-oxide semiconductors (or simply CMOS devices) have an
important role in the development of the information and electronic
industry. A scheme of this device can be represented basically by figure \ref%
{transistor metal-oxide}. As can be observed, it is composed by a thin metal
plate, followed by an insulating layer (e.g., SiO$_{2}$), and finally a
semiconductor layer (Silicon-Si). The working of device supposes two
voltages:

\begin{enumerate}
\item The gate-source voltage ($V_{GS}$), which controls the Fermi-level of
charge carriers in the semiconductor, i.e., the electrons are attracted with
larger or smaller intensity to the interface between the oxide and
semiconductor according to the magnitude of $V_{GS}$;

\item The drain source voltage ($V_{DS}$), responsible to move the electrons
to compose the current (drain current). The uniformity of this current can
be affected by charge traps found close to the semiconductor-insulator
interface. In this context the rules of capture and emission of charge
carriers by traps in semiconductor devices lead to irregular signals (noise)
in the current which can be observed in figure \ref{rts}.
\end{enumerate}

In this figure we observe how the current of device is affected by a single
trap. The fluctuations experimentally observed are caused by the effect of
the superposition of a number of traps, under different conditions and
moreover, we can consider fluctuations from a sample of devices for
completely describing this phenomena (see for example \cite{silva2006}).

Such fluctuations are called random telegraph signals \cite{Machlup1954}. In
other contexts, as for example Ni/NiO/Co junctions \cite{Doudin1997}, such
fluctuations are even able to govern magnetoresistance and their study also
will be very important in the context of nanostructures (see for example
\cite{Skocpol1986}, \cite{Dekker1991}).

The capture and emission of charge carriers by the traps is described as a
simple Poisson process governed by rates $\tau _{c}$ and $\tau _{e}$, where
the capture occurs with probability $p(0\rightarrow 1)dt=\tau _{c}^{-1}dt$
and and emission $p(1\rightarrow 0)dt=\tau _{e}^{-1}dt$

\begin{figure}[tbp]
\begin{center}
\includegraphics[width=0.9\columnwidth]{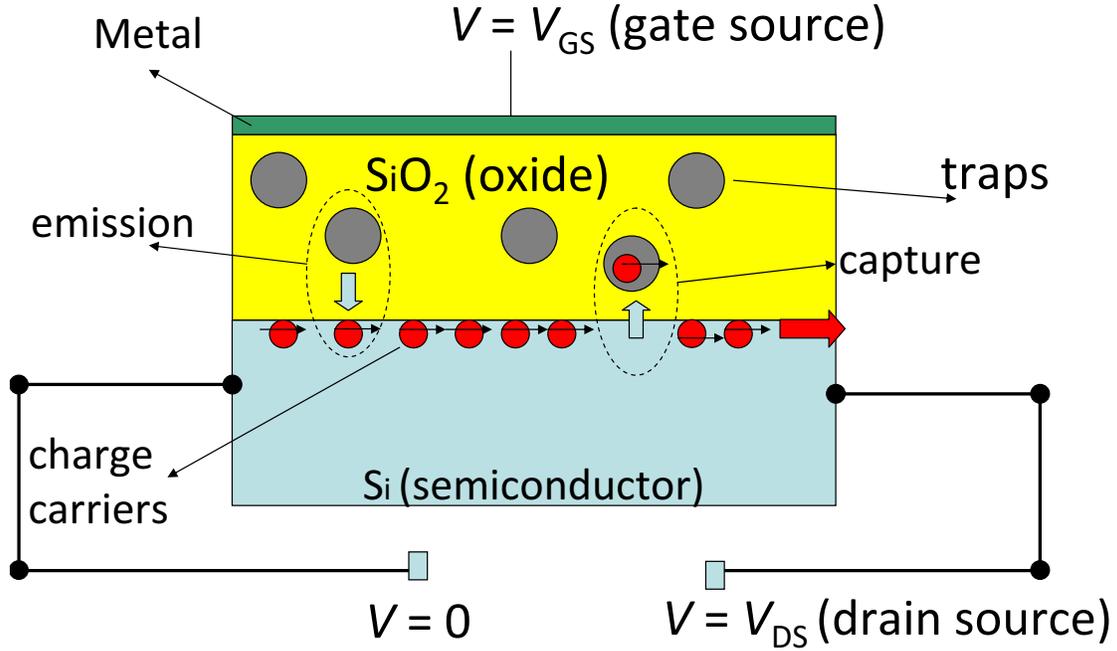}
\end{center}
\caption{A scheme of a complementary metal oxide semiconductor device (CMOS
transistor), composed by a fine metal plate, an oxide (SiO{\protect\small 2}%
), and a semiconductor (Si). A voltage $V_{GS}$ (gate source) has a role of
attracting the electrons to the top of semiconductor near of the oxide.
Other voltage $V_{DS}$ moves the charge carriers generating a drain current.
The charge carriers under a Fermi Level can be captured or emitted come back
to the current along time evolving according to Fermi-Dirac statistics.
These successive captures and emissions generate the known random telegraph
signals. }
\label{transistor metal-oxide}
\end{figure}

Such rates $\tau _{c}$ and $\tau _{e}\ $can be defined as the time average
in state 1 and state 0 respectively: $\tau _{c}=\left\langle t\right\rangle
_{1}=\int_{0}^{\infty }t\ P_{1}(t)dt$ and $\tau _{e}=\left\langle
t\right\rangle _{0}=\int_{0}^{\infty }t\ P_{0}(t)dt$, where $P_{1}(t)$ is
the probability of permanency in state 1 and $P_{0}(t)$ the respective
amount for state 0. Naturally $P_{1}(t)=1/\tau _{c}\cdot e^{-t/\tau _{c}}$
and $P_{0}(t)=1/\tau _{e}\cdot e^{-t/\tau _{e}}$.

If the number of trapped charge carriers increases over time, a decrease of
the current may be observed. This is an aging effect usually called bias
temperature instability (n- or p-bti), since it depends on bias (Fermi
level) and temperature, as discussed below.

In this context the degradation of a MOS transistor can be measured as the
number of occupied traps and the dynamics of this occupation must be better
understood. In this paper we aim at a theoretical analysis to describe the
density of occupied traps in a semiconductor device and so to understand how
a characteristic degradation process occurs in these devices and other
similar devices.

So, first of all, we need to calculate the probability of a particular trap
with constants $\tau _{c}$ and $\tau _{e}$ starting from state $0$ (empty)
and after a elapsed time $t$ it is in this same state, which we denote $%
p_{00}(t)$. This probability can be calculated observing that (\cite%
{Kirton1989}, \cite{Machlup1954}):
\begin{equation*}
P_{01}(t+dt)=P_{01}(t)p(1\rightarrow 1)+P_{00}(t)p(0\rightarrow 1)
\end{equation*}%
where $p(0\rightarrow 1)=dt/\tau _{c}$ and $\ p(1\rightarrow
1)=1-p(1\rightarrow 0)=1-dt/\tau _{e}$ and also $P_{00}(t)=1-P_{01}(t)$.
This leads to a simple differencial equation:$\ dP_{01}(t)/dt=\tau
_{c}^{-1}-(\tau _{e}^{-1}+\tau _{c}^{-1})P_{01}(t)$. If $P_{00}(0)=1$, its
solution is $P_{01}(t)=\tau _{e}(\tau _{e}+\tau _{c})^{-1}\left[ 1-\exp
(-t/\tau _{eq})\right] $, where $1/\tau _{eq}=1/\tau _{e}+1/\tau _{c}$.
Similar results can be performed leading to $P_{11}(t)=\dfrac{\tau _{e}}{%
\tau _{e}+\tau _{c}}\left[ \tau _{e}+\tau _{c}\exp (-t/\tau _{eq})\right] $.
Unless fluctuations on the current amplitude, whose average $\Delta $
depends on other microscopic factors of the device, this probability
corresponds to the autocorrelation of the system $A(t)=\left\langle \sigma
(0)\sigma (t)\right\rangle $, where $\sigma (t)$ corresponds to the state of
the trap (occupied $\sigma =0$ or empty $\sigma =1)$. In frequency domain
this exponential decay of autocorrelation for one trap is described by
Lorentzians, since the power spectrum density, i.e., the fourier transform
of the autocorrelation is
\begin{equation*}
S(f)=\int_{-\infty }^{\infty }e^{-2\pi fti}A(t)dt=\frac{4\Delta ^{2}}{(\tau
_{e}+\tau _{c})(\tau _{eq}^{2}+2\pi f^{2})}
\end{equation*}%
The known 1/$f$ noise results from a sum of these Lorentzians (a
contribution of the many traps in device). For more details about the origin
of 1/$f$ noise, see for example \cite{Kaulakys1998},\cite{silva2006}, \cite%
{Ralls1984}, \cite{Rogers1984}.

So, coming back to relaxation phenomena, and starting from $n(0)=0$ (all
traps empty), we can calculate the average density of occupied traps at time
$t$%
\begin{equation}
\begin{array}{lll}
\left\langle n(t)\right\rangle & = & \left\langle \sum_{k=0}^{N_{tr}}\sigma
_{i}(t)\right\rangle \\
&  &  \\
& = & \sum_{k=0}^{N_{tr}}k\Pr (k|n(0),t)%
\end{array}
\label{average_density_1}
\end{equation}%
where $\Pr (k|n(0),t)$ is the probability of just $k$ traps are occupied at
time $t$, with $k=0...N_{tr}$. But the traps have different constants $\tau
_{e}$ and $\tau _{c}$ and from that, we write%
\begin{equation*}
\Pr (k|n(0),t)=\sum_{C_{k}}\prod\limits_{i=1}^{k}P_{01}(\tau
_{c}^{(d_{i})},\tau
_{e}^{(d_{i})};t)\prod\limits_{i=k+1}^{N_{tr}}P_{00}(\tau
_{c}^{(d_{i})},\tau _{e}^{(d_{i})};t)
\end{equation*}%
with $C_{k}$ denoting every subset $\{d_{1},d_{2},...,d_{k}\}$ from $%
\{1,2,...,,n\}$. But $\left\{ \tau _{c}^{(d_{i})},\tau
_{e}^{(d_{i})}\right\} _{i=1}^{N_{tr}}$ are statistically independent and
identically distributed, and we have :

\begin{equation*}
\begin{array}{lll}
\overline{\Pr (k|n(0),t)} & = & \sum_{C_{k}}\overline{P_{01}(\tau _{c},\tau
_{e};t)}^{k}\cdot \ \ \overline{P_{00}(\tau _{c},\tau _{e};t)}^{N_{tr}-k} \\
&  &  \\
& = & \binom{n}{k}\overline{P_{01}(\tau _{c},\tau _{e};t)}^{k}\cdot \ \
\overline{P_{00}(\tau _{c},\tau _{e};t)}^{N_{tr}-k}%
\end{array}%
\end{equation*}%
where $\overline{\cdot }=\int \int \cdot \ f(\tau _{c})g(\tau _{e})d\tau
_{c}d\tau _{e}$, where $f(\tau _{c})$ and $g(\tau _{e})$ are probability
densities of time constant times of capture and emission. Microscopically
these quantities can be better understood. Actually, $\tau _{c}$ and $\tau
_{e}$ have a thermal and a gate voltage dependence. Some more detailed
approaches use quantum two-dimensional calculations of these quantities \cite%
{Palma1997}. Here we use a known simplification proposed by Kirton and Uren
\cite{Kirton1989}, where $\tau _{c}$ and $\tau _{e}$ are random variables
that follow the form: $\tau _{c}=10^{p}(1+\exp (-q))$ and $\tau
_{e}=10^{p}(1+\exp (q))$, where $p\in \lbrack p_{\min },p_{\max }]$ and$\
q=(E_{t}-E_{F})/k_{B}T\in \lbrack E_{v}-E_{F})/k_{B}T,(E_{c}-E_{F})/k_{B}T]$
are randomly distributed according to respectively a uniform and a u-shape
distribution.

There is no much information about these density of states of the traps in
literature, but Wong and Cheng \cite{Wong1990} show that for 3 different
prepared gate oxides it follows a u-shape form.

Naturally, we must observe that $\tau _{eq}=\tau _{c}\tau _{e}/(\tau
_{c}+\tau _{e})=10^{p}$ corresponds to an uniform distribution of time
constants ($\tau _{eq}$) in a log scale, as expected. Here, $E_{t}$ and $%
E_{F}$ are respectively the energy of observed trap and Fermi Level of
system that is directly proportional to $V_{GS}$ applied in device. For our
purpose, it is more interesting to switch our average:%
\begin{equation*}
\int \int \left( \cdot \right) \ f(\tau _{c})g(\tau _{e})d\tau _{c}d\tau
_{e}\rightarrow \frac{\int\nolimits_{p_{\min }}^{p_{\max
}}\int\nolimits_{E_{v}}^{E_{c}}\left( \cdot \right) \ dp\ \Omega
(E_{t})dE_{t}}{\left( \int_{E_{v}}^{E_{c}}\ \Omega (E_{t})dE_{t}\right)
(p_{\max }-p_{\min })}
\end{equation*}%
where $\Omega (E_{t})$ is the density of states of the traps in the
interface.

\ Coming back to equation \ref{average_density_1}, after some straithforward
calculations we have

\begin{equation*}
\begin{array}{lll}
\overline{\left\langle n(t)\right\rangle } & = & \sum_{k=0}^{N_{tr}}k\binom{n%
}{k}\overline{P_{01}(p,E_{t};t)}^{k}\cdot \ \ \overline{P_{01}(p,E_{t};t)}%
^{N_{tr}-k} \\
& = & N_{tr}\overline{P_{01}(p,E_{t};t)} \\
& = & N_{tr}\int\nolimits_{E_{v}}^{E_{c}}\frac{dE_{t}\ \Omega (E_{t})}{%
1+e^{-(E_{t}-E_{F})/k_{B}T}}dE_{t}\cdot \\
&  & \cdot \frac{1}{(p_{\max }-p_{\min })}\int_{p_{\min }}^{p_{\max }}dp%
\left[ 1-\exp (-10^{-p}t)\right]%
\end{array}%
\end{equation*}

The second integral must be better worked out. Making a suitable change
variable $p=-\log _{10}$ $(\frac{u}{t})$, $dp=-\ln ^{-1}10\frac{du}{u}$ and
we have the temperature dependence separated of time dependence via two
integrals:
\begin{align}
\overline{\left\langle n(t)\right\rangle }& =N_{tr}\left(
\int\nolimits_{E_{v}}^{E_{c}}\frac{dE_{t}\ \Omega (E_{t})}{%
1+e^{-(E_{t}-E_{F})/k_{B}T}}\right) \cdot  \label{many_traps} \\
& \cdot \left[ \frac{\ln ^{-1}10}{(p_{\max }-p_{\min })}\int_{10^{-p_{\min
}}t}^{10^{-p\max }t}du\frac{\left( e^{-u}-1\right) }{u}\right]
\end{align}%
which can be analyzed numerically. A particular case is when $f(E_{t})$ is
uniform, and in this case we have $\int\nolimits_{E_{v}}^{E_{c}}\frac{%
dE_{t}\ \Omega (E_{t})}{1+e^{-(E_{t}-E_{F})/k_{B}T}}=\frac{k_{B}T}{%
(E_{c}-E_{v})}\ln \left[ \frac{%
e^{(E_{c}-E_{v})/k_{B}T}+e^{(E_{F}-E_{v})/k_{B}T}}{1+e^{(E_{F}\
-E_{c})/k_{B}T}}\right] $. If $E_{F}$ $=(E_{v}+E_{c})/2$, i.e., it is
exactly in the middle of band gap, this integral is numerically equal $1/2$
and there is no temperature dependence, i.e.,

\begin{equation}
\overline{\left\langle n(t)\right\rangle _{uniform}}=\frac{N_{tr}\ln^{-1}10}{%
2(p_{\max}-p_{\min})}\int_{10^{-p_{\min}}t}^{10^{-p\max}t}du\frac{\left(
e^{-u}-1\right) }{u}  \label{many_traps_uniform}
\end{equation}

A simple particular case is if we observe the evolution \ of the occupation
probability of a single trap, with time constants $\tau _{c}$ and $\tau _{e}$%
, is numerically \ equal to $\lim_{N_{tr}\rightarrow \infty }$ $\overline{%
\left\langle n(t)\right\rangle _{uniform}}/N_{tr}$ when all traps have the
same $\tau _{e}$ and $\tau _{c}$, which is given by:
\begin{equation}
\Pr (\sigma _{i}(t)=1)=\dfrac{1-\exp \left( -\tfrac{1+\beta }{\beta \ \tau
_{e}}t\right) }{1+\beta }  \label{one_trap}
\end{equation}%
where $\beta =\tau _{c}/\tau _{e}$ is a important ratio considered in this
context. Naturally $\Pr (\sigma _{i}(t)=1)\rightarrow $ $\frac{1}{1+\beta }=%
\frac{\tau _{e}}{\tau _{e}+\tau _{e}}$ when $t\rightarrow \infty $. This
leads to a simple but important conclusion: if $\beta >1$ (time capture is
greater than time emission) we have $\Pr (\sigma _{i}=1)<1/2$ and otherwise
-- i.e., $\beta <1$(time emission greater than time capture), $\Pr (\sigma
_{i}=1)>1/2$ when $t\rightarrow \infty $.

But what is the behavior of $\overline{\left\langle n(t)\right\rangle }$ in
a realistic case(i.e., when $\tau _{c}$ and $\tau _{e}$ are randomly
distributed)? In this case we solve numerically the exponential integral
from \ref{many_traps}. We adopt usual values found in the literature for
this problem ($p_{\min }=0$ and $p_{\max }=7$) what means a frequency
ranging from 1 to 10$^{7}$ Hz.

\begin{figure}[tbp]
\begin{center}
\includegraphics[width=0.9\columnwidth]{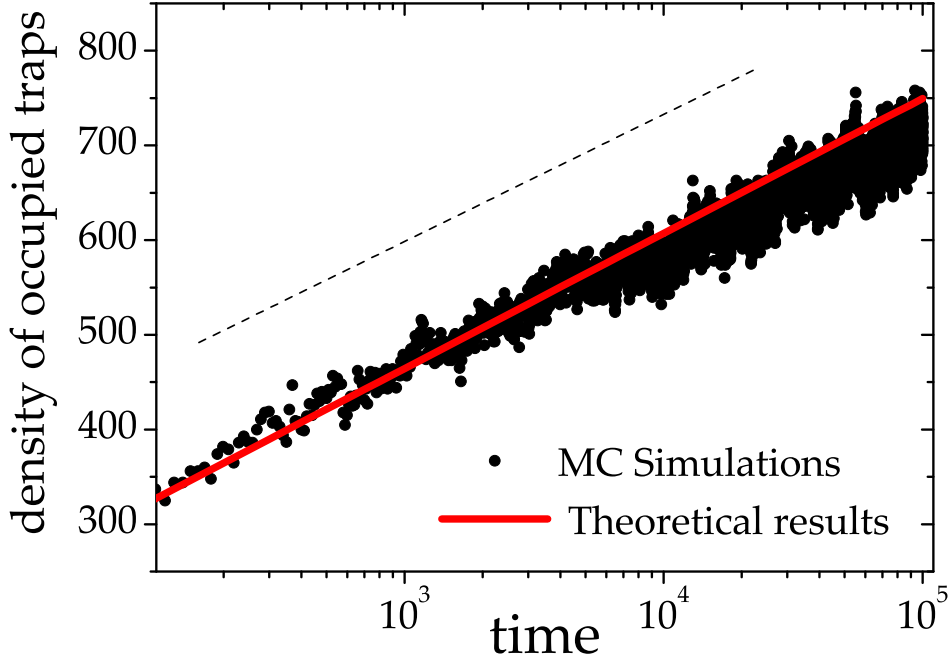}
\end{center}
\caption{Time evolving of $\overline{\left\langle n(t)\right\rangle }$
obtained directly by numerical integration of equation \protect\ref%
{many_traps_uniform}.The points corresponds to MC simulations performed in
same conditions. }
\label{density_of_traps_many}
\end{figure}

The continuous curve in figure \ref{density_of_traps_many} shows the time
evolving $\overline{\left\langle n(t)\right\rangle }$ theoretically obtained
(i.e. numerical integration of equation \ref{many_traps}).

The points corresponds to our MC simulations. For these MC simulations, we
start from a given number of empty traps. For each time, each empty trap $%
i=1,...,N_{tr}$ becomes occupied with probability $p_{i}(0\rightarrow
1)=10^{-p_{i}}(1+e^{-q_{i}})^{-1}$ and similarly an occupied trap becomes
empty again with probability $p_{i}(1\rightarrow
0)=10^{-p_{i}}(1+e^{q_{i}})^{-1}$, where $p_{i}$ is uniformly drawn in $%
[p_{\min },p_{\max }]$, and here the same values ($p_{\min }=0$ and $p_{\max
}=7$) were used. Similarly $E_{t}^{(i)}$is uniformly drawn in $[E_{v},E_{c}]$%
. For real devices $E_{t}$ should change from $\ \symbol{126}\ 0.2\ $eV to $%
\ \symbol{126}\ 1.0$ eV. First of all, $E_{F}$ corresponds to middle of
band-gap ($E_{F}\ \symbol{126}\ 0.6\ $eV$)$, $k_{B}=8.617385\times 10^{-5}$%
eV K$^{-1}$and $T=300$ K, which leads to $q_{i}\in \lbrack -15.473,15.473]$.
We can observe a excellent agreement between the MC simulations and our
theoretical equations, showing that in semi-log plot in figure \ref%
{density_of_traps_many} the relaxation dynamics follows a logarithm law:%
\begin{equation*}
\overline{\left\langle n(t)\right\rangle _{uniform}}\ \sim \ A+B\log t
\end{equation*}%
where we find $A=31.80(15)$ and $B=144.20(3)$. The uncertainties were
obtained, using error bars obtained from 16 independent runs of the program.
We test other temperature values but we did not find difference as expected
when $E_{F}$ is in the middle of band gap in this case where a uniform
density of states is considered.

Extending our results, we can compute the second moment:

\begin{equation*}
\overline{\left\langle n(t)^{2}\right\rangle }=\sum_{k=0}^{N_{tr}}k^{2}%
\overline{\Pr(k|n(0),t)}
\end{equation*}
which yields $\overline{\left\langle n(t)^{2}\right\rangle }=N_{tr}\overline{%
\Pr(k|n(0),t)}+N_{tr}(N_{tr}-1)\overline{\Pr(k|n(0),t)}^{2}\approx\overline{%
\left\langle n(t)\right\rangle }+\overline{\left\langle n(t)\right\rangle }%
^{2}$

So we have $\left\langle n(t)^{2}\right\rangle \approx \frac{N_{tr}\ln
^{-1}10}{2(p_{\max }-p_{\min })}\int\nolimits_{10^{-p_{\min }}t}^{10^{-p\max
}t}du\frac{\left( e^{-u}-1\right) }{u}+\frac{\ln ^{-2}10N_{tr}^{2}}{%
4(p_{\max }-p_{\min })^{2}}\left[ \int\nolimits_{10^{-p_{\min
}}t}^{10^{-p\max }t}du\frac{\left( e^{-u}-1\right) }{u}\right] ^{2}$, which
leads to

\begin{equation*}
\overline{\left\langle n(t)^{2}\right\rangle }\sim A(1+A)+B(1+2A)\log
t+B^{2}\log ^{2}t
\end{equation*}%
with $A$ and $B$ exactly as reported before.

\begin{figure}[tbp]
\begin{center}
\includegraphics[width=0.9\columnwidth]{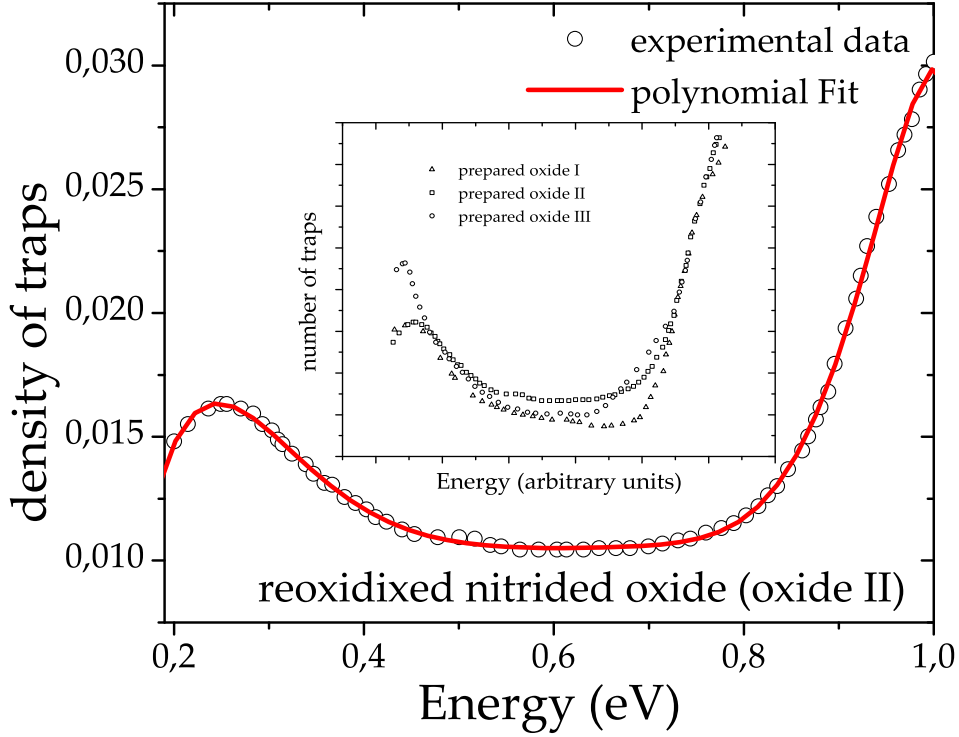}
\end{center}
\caption{A polynomial fit of a u-shape (Reoxidized Nitrided-oxide). The
experimental curves were extracted from \protect\cite{Wong1990}. The inside
plot corresponds to the other u-shape densities that are very similar. }
\label{u-shape}
\end{figure}

Looking at the temperature dependence, we must analyze more realistic
densities of states. After a detailed scanning of the plots for the
densities of states, for the 3 prepared gate oxides (TCE Oxide, Reoxidized
Nitrided-oxide, and Nitrided-oxide) found in the reference \cite{Wong1990},
a fitting by a eighty-degree polynomial, here described by $\widehat{\Omega }%
(E_{t})=\sum\nolimits_{k=0}^{8}\beta _{k}E_{t}^{k}$ were performed (see \cite%
{Silva2008} for a more detailed discution of this part). This excelent fit
can be seen in figure \ref{u-shape}.

Using these u-shaped densities of states or even their polynomial fit, we
can calculate the temperature dependence:

\begin{equation}
\varphi (T,E_{F})=\frac{1}{Z}\sum\nolimits_{k=0}^{8}\beta
_{k}\int\nolimits_{E_{v}}^{E_{c}}\frac{E_{t}^{k}dE_{t}\ }{%
1+e^{-(E_{t}-E_{F})/k_{B}T}}  \label{equation_fi}
\end{equation}%
with $Z=\sum\nolimits_{k=0}^{8}\beta _{k}(k+1)^{-1}(E_{c}^{k+1}-E_{v}^{k+1})$
where then $\overline{\left\langle n(t)\right\rangle }$ $\sim \ \varphi
(T,E_{F})(A+B\log t),$ showing that temperature and Fermi level dependence
are independent of time, i.e., the relaxation depends logarithmically on
time for each fixed pair ($T,E_{F}$).

\begin{figure}[tbp]
\begin{center}
\includegraphics[width=0.9\columnwidth]{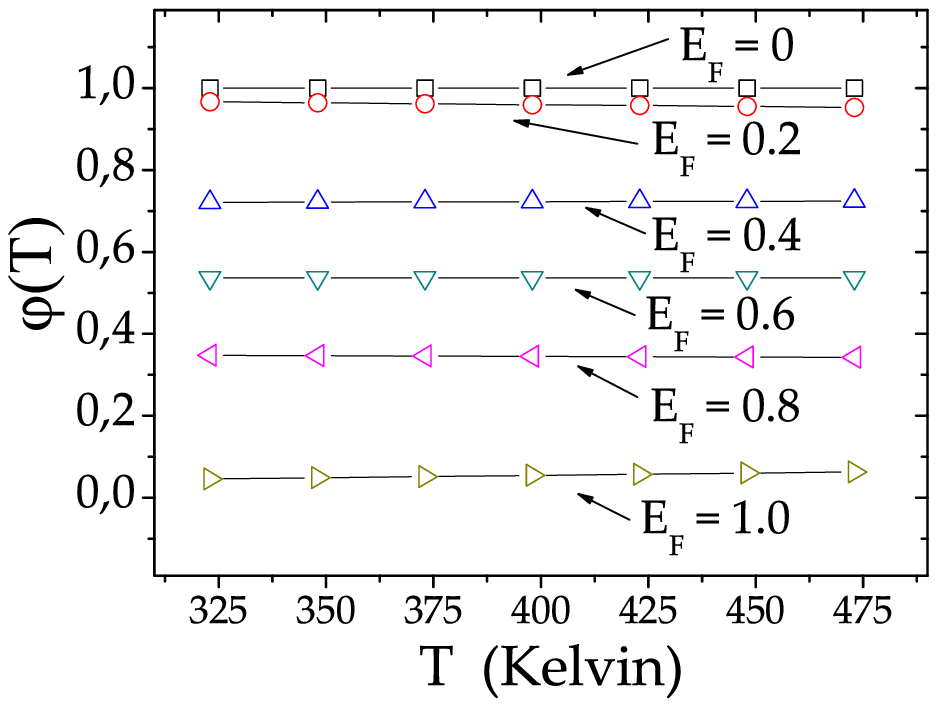}
\end{center}
\caption{Fermi level dependence of $\protect\varphi (T)$. }
\label{femi_level_dependence}
\end{figure}

Using Simpson numerical integration, we calculate $\varphi (T)$ and now the
time evolving of $\overline{\left\langle n(t)\right\rangle }$ is calculated
according to equation \ref{many_traps}. By motivations from experimental
results, initially we study $\overline{\left\langle n(t)\right\rangle }$ for
a Fermi level set on $E_{F}=Ec=1$eV using the u-shape obtained from
Reoxidized Nitrided-oxide.
\begin{figure}[tbp]
\begin{center}
\includegraphics[width=0.9\columnwidth]{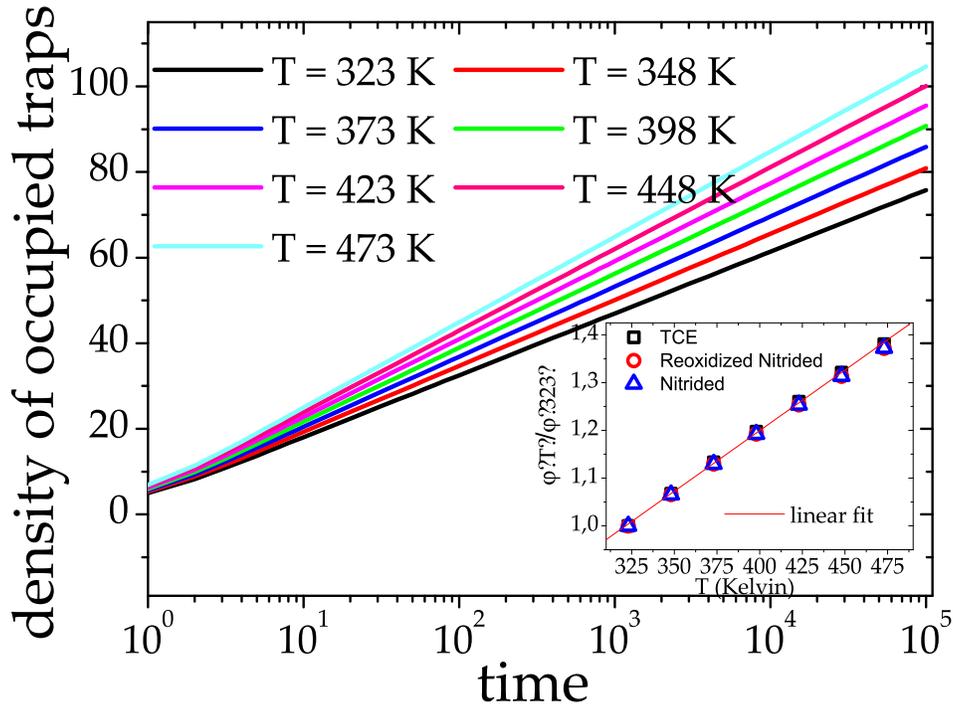}
\end{center}
\caption{Density of occupied traps for different temperatures ($323$, $373$,
$423$, $448$, $473$ $K$) using a real u-shape, we observe effects on
logarithmic law exactly as observed via experimental results obtained in
\protect\cite{Tibor2009}. The inside plot shows the amplification $\protect%
\varphi (T)/\protect\varphi (323)$ as function of temperature for the 3
different oxides showing a agreement among them indicating a linear
universal behavior $\protect\varphi (T)/\protect\varphi (323)\sim 0.18T$. }
\label{density_u-shape}
\end{figure}
The figure \ref{density_u-shape} shows $\overline{\left\langle
n(t)\right\rangle }$ for different temperatures. The inside plot in this
same figure shows the amplification factor of temperature $A(t)=\varphi
(T)/\varphi (323)$, since $T_{0}=323$K\ ($50^{0}$C) is the minimum
temperature used in our calculations, for the 3 different oxides extracted
from \cite{Wong1990}. We show the linear universal behavior, described by
relation
\begin{equation*}
A(T)\sim 0.18\ T
\end{equation*}

\ It is also interesting to study the dependence on Fermi level. The Fermi
level may be varied by changing the gate bias of the device as
experimentally explored in \cite{Tibor2009}. We studied the dependence of $%
\varphi (T)$ as function of $T$ for different values of $E_{F}$. Our results
show that the curves showing the evolution of relaxation as function of time
can be collapsed in a single one if multiplied by a suitable constant. This
constant is $\varphi (T,E_{F})$ (see figure \ref{femi_level_dependence}).
Hence, our results are in agreement to the experimental findings for the
Fermi-level dependence of the relaxation.

\begin{figure}[tbp]
\begin{center}
\includegraphics[width=0.9\columnwidth]{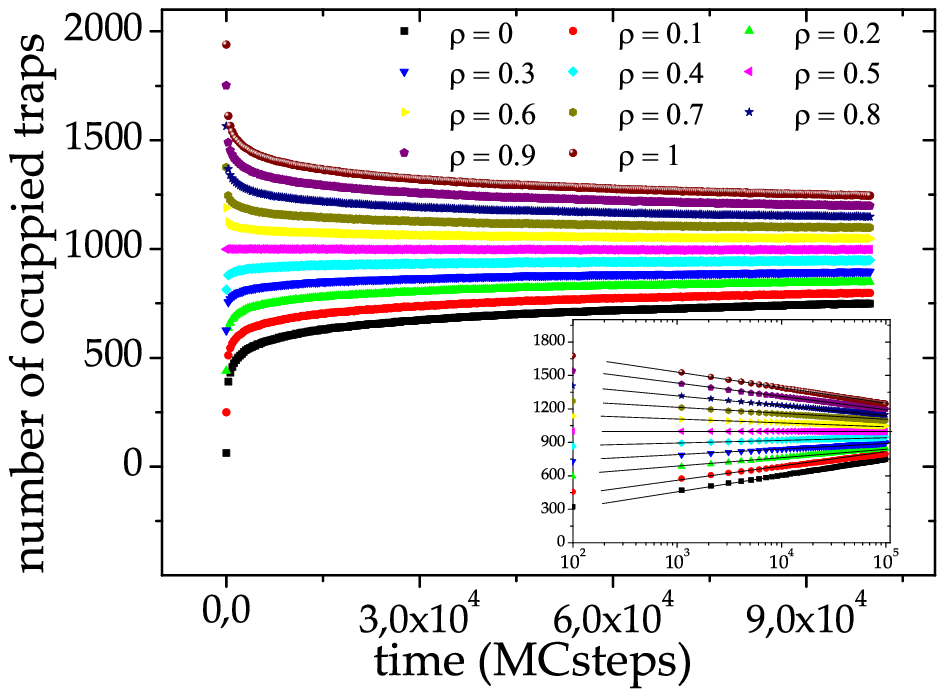}
\end{center}
\caption{The density of states increases logarithmically for $\protect\rho %
_{0}<0.5$, and decreases for $\protect\rho _{0}>0.5$. For $\protect\rho %
_{0}=0.5$ we can see a constant behavior of degradation, i.e., the number of
occupied traps keeps inalterated along time. The inside plot is a semi-log
plot, showing the logarithmic behavior }
\label{initial_effects}
\end{figure}

We did also study the dependence on the initial density of occupied traps.
In this case, if the density of initially occupied traps is lower than the
equilibrium value, in the relaxation process, the number of occupied traps
increases logarithmically. On the other hand, if the density of initially
occupied traps is higher than the equilibrium value, the density of occupied
states decreases logarithmically in the relaxation process. MC simulation
did confirm this behavior. MC simulations were performed starting from
different initial density of occupied traps $\rho _{0}=0,\ 0.1,\ 0.2,\ 0.3,\
...,\ 1.0$, where the equilibrium value is $\rho =0.5$. We can observe this
behavior in figure \ref{initial_effects}.

In summary our results corroborate the experimentally observed logarithmic
relaxation of the density of the occupied traps via MC simulations and from
theoretical analysis in complementary metal-oxide semiconductors governed by
Fermi-Dirac-Shockley-Read Statistics \cite{Shockley1952}. Our results also
corroborate the experimentally observed temperature dependence, which shows
that the relaxation as a function of time at different temperatures may be
collapsed into a single curve using a suitable scaling factor. This behavior
is experimentally observed e.g. in \cite{Tibor2009}. The scaling factor is
the function $\varphi (T,E_{F})$ of equation \ref{equation_fi}.

\end{document}